\documentclass[prd,superscriptaddress,a4paper,10pt,nofootinbib]{revtex4}
\usepackage{graphicx}
\usepackage{graphics}
\usepackage{epsfig}
\usepackage{dcolumn}
\usepackage{bm}
\usepackage{multirow}
\usepackage{tabularx}
\usepackage{hyperref}
\usepackage{commath}
\usepackage{epstopdf}
\usepackage[T1]{fontenc}
\usepackage{geometry}
\usepackage{color}
\usepackage{threeparttable}

\newcommand{\epem}{\mathrm{e}^+\mathrm{e}^-}
\newcommand{\gaga}{\gamma\gamma}
\newcommand{\delphes}{\textsc{Delphes}}
\newcommand{\LumiInt}{\mathcal{L}_\text{int}}
\newcommand{\sqrts}{\sqrt{s}}
\newcommand{\abinv}{ab$^{-1}$}

\begin{document}
\title{Searches for axion-like particles via $\gaga$ fusion at future $\epem$ colliders}

\author{Patricia Rebello Teles}
\email{patricia.rebello.teles@cern.ch}
\affiliation{Centro Brasileiro de Pesquisas Físicas (CBPF), Rua Dr.\ Xavier Sigaud, 150 Urca, 22290-180, Rio de Janeiro, RJ, Brazil. \\
}

\author{David d'Enterria}
\email{david.d'enterria@cern.ch}
\affiliation{CERN, EP Department 1211 Geneva, Switzerland\\
}

\author{Victor P. { Gon\c{c}alves}}
\email{barros@ufpel.edu.br}
\affiliation{Institute of Physics and Mathematics, Federal University of Pelotas, \\
Postal Code 354,  96010-900, Pelotas, RS, Brazil}
\affiliation{Institute of Modern Physics, Chinese Academy of Sciences,
Lanzhou 730000, China}

\author{Daniel E. Martins}
\email{daniel.ernani@ifj.edu.pl}
\affiliation{The Henryk Niewodniczanski Institute of Nuclear Physics (IFJ)\\ Polish Academy of Sciences (PAN), 31-342, Krakow, Poland
}

\begin{abstract}
Opportunities for searches for axion-like particles (ALPs) coupling to photons in $\epem$ collisions at the Future Circular Collider (FCC-ee) and International Linear Collider (ILC) are investigated. We perform a study of the photon-fusion production of ALPs decaying into two photons, $\epem \overset{\gaga}{\longrightarrow} \mathrm{e}^{+}\;a(\gaga)\;\mathrm{e}^{-}$, over the light-by-light continuum background, for the planned FCC-ee and ILC center-of-mass energies and integrated luminosities. An analysis of the feasibility measurements is presented using parametrized simulations for two types of detectors. Upper limits at 95\% confidence level (CL) on the cross section for ALP production, $\sigma(\gaga \to a \to \gaga)$, and on the ALP-photon coupling are obtained over the $m_a \approx 0.1$--1000~GeV ALP mass range, and compared to current and future collider searches. Production cross sections down to $\sigma(\gaga \to a \to \gaga) \approx 1$~fb~(1~ab) will be probed at $m_a\approx 1$~(300)~GeV, corresponding to constraints on the axion-photon coupling as low as $g_\mathrm{a\gaga} \approx 2\cdot10^{-3}$~TeV$^{-1}$.
\end{abstract}



\maketitle

\section{Introduction}

After more than a decade of operation of the Large Hadron Collider (LHC) with proton-proton collisions at the energy frontier, no new elementary particle beyond the Higgs boson~\cite{ATLAS:2012yve,CMS:2012qbp} has been observed. Aside from the scalar boson discovery~\cite{Bass:2021acr}, the main fundamental questions that motivated the LHC construction remain therefore open~\cite{Bose:2022obr}: What is dark matter (DM)? What is the origin of matter-antimatter asymmetry in the universe? How are the neutrino masses generated? What physical mechanism protects the mass of the scalar Higgs field from quantum corrections up to the Planck scale  without a high degree of fine tuning? Why there is no visible violation of charge-parity symmetry (CP) in the strong interaction? What is the origin of the seemingly arbitrary structure of fermion masses (Yukawa couplings) and mixings?
Hierarchy arguments based on the naturalness of the electroweak (EW) scale suggest that new physics degrees of freedom may exist at or below the TeV scale~\cite{Giudice:2008bi}. Such expectations are now in tension with null results from new physics searches at the LHC, which imply either a significant mass gap between the EW scale and the scale of physics beyond the Standard Model (BSM) that stabilizes it, or new physics states that are light, but very weakly coupled. In many scenarios, these new light physics states are the (pseudo)Nambu--Goldstone bosons that arise from spontaneously breaking a high-energy global symmetry or in low-energy effective theory of string compactifications. In this context, in parallel to searches for new phenomena at high masses and transverse momenta, studies at current and future colliders have broadened their scope to include also searches for new particles with very suppressed interactions with the Standard Model (SM) bosons and/or fermions~\cite{Agrawal:2021dbo}. Among such feebly-interacting particles (FIPs), generic pseudoscalar axion-like particles (ALPs) are a prime bosonic DM candidate~\cite{Dine:1982ah,Abbott:1982af,Preskill:1982cy,Duffy:2009ig}, can provide an elegant solution to the strong CP~\cite{Peccei:1977hh} and/or hierarchy~\cite{Graham:2015ouw} problems, and are also ubiquitous in string theory realizations~\cite{Ringwald:2012cu}.

In order to explore the broad range in mass and couplings that ALP physics suggests~\cite{Irastorza:2021tdu}, many experiments and techniques have been developed. Heavy ALPs with masses above $m_a \approx 0.1$~GeV can be searched for at current~\cite{Mimasu:2014nea,Knapen:2016moh,Brivio:2017ije,Bauer:2017ris,Dolan:2017osp,CMS:2018erd,Aloni:2019ruo,Belle-II:2020jti,ATLAS:2020hii,dEnterria:2021ljz,TOTEM:2021zxa,CMS:2022zfd,ATLAS:2023zfc,Cerci:2021nlb,BuarqueFranzosi:2021kky,BESIII:2022rzz,Harland-Lang:2022jwn} and future~\cite{Bauer:2018uxu, Yue:2021iiu,Steinberg:2021wbs,Han:2022mzp,Wang:2022ock,Inan:2022rcr,Tian:2022rsi,Cheung:2023nzg,Mosala:2023sse,Balkin:2023gya} high-energy colliders. 
The particle physics priority for a post-LHC machine is an electron-positron ($\epem$) facility~\cite{EuropeanStrategyforParticlePhysicsPreparatoryGroup:2019qin,Snowmass2022}, such as the Future Circular Collider (FCC-ee)~\cite{FCC:2018byv,FCC:2018evy} or the International Linear Collider (ILC)~\cite{Behnke:2013xla}, aiming at very precisely probing the Higgs sector of the SM. Such machines have also the capability to accurately search for new FIPs in low mass/coupling regions inaccessible at the LHC. While new particles with 0.1 to 100~GeV masses are difficult to access at hadron colliders due to trigger limitations and large backgrounds in p-p collisions ---despite photon-fusion processes in ultraperipheral collisions of heavy ions~\cite{Baltz:2007kq} providing somewhat more favorable conditions in this mass range~\cite{Bruce:2018yzs,Goncalves:2021pdc,dEnterria:2022sut,Shao:2022cly}--- the clean environment and large integrated luminosities available 
at the FCC-ee and ILC lepton colliders will render them very sensitive machines in searches for EW-coupled ALPs.\\

\begin{figure}[htpb!]
\centering
\includegraphics[width=0.99\textwidth]{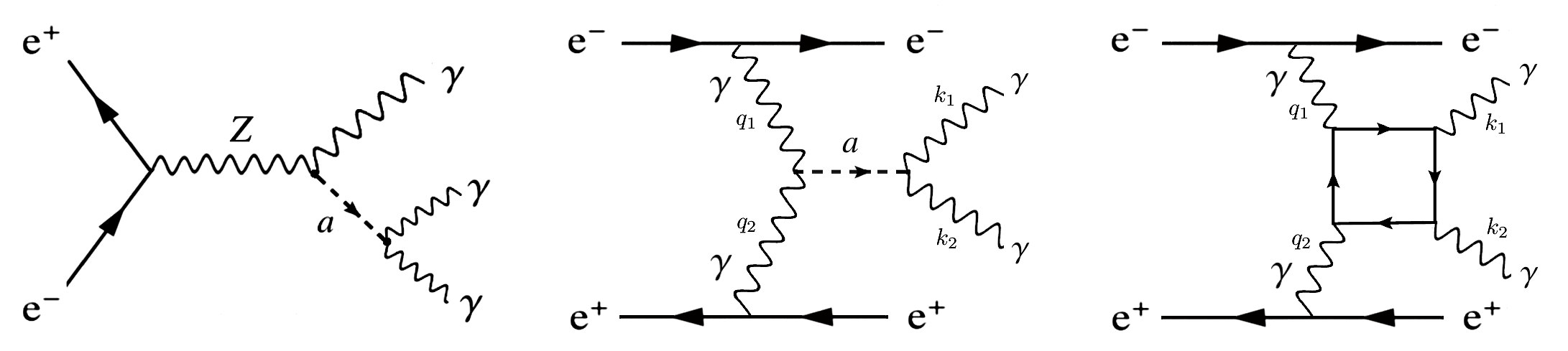}
\caption{Schematic diagrams of ALPs production in $\epem$ collisions via Z boson decays (left) and in $\gaga$ fusion (center), and light-by-light background of the latter process (right). In the photon-fusion diagrams, the incoming (outgoing) photons have momenta $q_{1,2}$ ($k_{1,2}$).}
\label{Fig:diagram_epem}
\end{figure}

In recent years, several projections for ALP search reaches have been presented for different future lepton colliders, including FCC-ee and ILC (see, e.g.~\cite{Bauer:2018uxu}) based on effective field theory models. In general, such studies have focused on ALPs coupling preferentially to photons (BC9 "photon dominance" benchmark point of Ref.~\cite{Agrawal:2021dbo}) and, more particularly, on
$\epem\to \mathrm{Z} \to a(\gaga)\,\gamma$ production (left diagram of Fig. \ref{Fig:diagram_epem}), which has vanishingly small backgrounds (the $\rm Z \to 3\gamma$ decay has a SM branching fraction of $\mathcal{B} = 0.85\cdot 10^{-9}$, including exclusive Z boson radiative meson decays~\cite{DdEDung:2023}), although at best, only approximate acceptances/efficiencies of the detectors have been taken into account. Alternatively, in this work we present an analysis of exclusive ALP production via photon fusion (Fig.~\ref{Fig:diagram_epem}, center) on top of the diphoton continuum from $\gaga$ interactions also known as light-by-light (LbL) scattering~\cite{dEnterria:2013zqi} (Fig.~\ref{Fig:diagram_epem}, right), taking into account realistic (fast) simulations of two typical future $\epem$ detectors~\cite{ILD:2019kmq,Antonello:2020tzq}. The ALPs event generation is performed with 
the  SuperChic~4.03~(SC4) Monte Carlo code~\cite{Harland-Lang:2020veo}, 
based on the equivalent photon approximation (EPA)~\cite{Budnev:1975poe}, accounting for the (small) virtualities of the emitted photons from the $\epem$ beams.
The generated events are further passed to parametrized responses of the FCC-ee and ILC detectors simulated using the \delphes-3 package~\cite{deFavereau:2013fsa}. For the FCC-ee, we consider the latest settings of the International Detector for Electron-positron Accelerator (IDEA) apparatus~\cite{Antonello:2020tzq}, and of the International Large Detector (ILD) concept~\cite{ILD:2019kmq} for the ILC case. The full analysis, from photon-photon scattering to the detector simulation is performed for varying values of ALP masses and photon-ALP couplings, using up-to-date FCC-ee and ILC integrated luminosities for their expected operation at various center-of-mass (c.m.) energies. We apply relevant event selection cuts in kinematic quantities reconstructed at the detector level, and show the regions of phase space with competitive sensitivities in the ALP mass versus $\gamma$-$a$ coupling plane. Preliminary results of this work have been presented elsewhere~\cite{WIN2021,BCLC2021}. This current work is the follow-up of our previous generator-level-only results.\\

The paper is organized as follows. In Section~\ref{sec:sim}, we present a brief review of the formalism needed to describe ALP and LbL production in $\gaga$ collisions at $\epem$ colliders. In particular, the EPA approach is reviewed and the Lagrangian used in the derivation of the Feynman rules relevant for the calculation of the photon-fusion ALP production is presented. In Section~\ref{sec:results}, the methodology of our analysis and the basic event selection criteria applied are discussed, and  the kinematic distributions for
ALP and LbL diphoton pairs produced in $\epem$ collisions at FCC-ee and ILC are presented. The expected upper limits on the cross sections and ALP-photon couplings are presented as a function of ALP mass for operation runs at different c.m.\ energies and luminosities, and a comparison with current upper bounds is also performed. Finally, the case where light ALPs are long-lived and can be identified by a macroscopically displaced diphoton vertex is discussed. A summary of our main results and conclusions is presented in Section \ref{sec:conclusions}.



\section{Simulation setup}
\label{sec:sim}

The rich physics possibilities of two-photon collisions were proposed long ago~\cite{Brodsky:1971ud,Budnev:1975poe} and have been an active topic of research at lepton~\cite{Vermaseren:1982cz,Uehara:1996bgt,Schuler:1997ex}, proton~\cite{deFavereaudeJeneret:2009db,dEnterria:2008puz,Harland-Lang:2015cta, Shao:2022cly}, and ion~\cite{Baur:2001jj,Baltz:2007kq,Shao:2022cly} colliders since. Future $\epem$ machines, in particular, will also provide an ideal environment to study $\gaga$ collisions at unprecedented energies and luminosities~\cite{RebelloTeles:2015oxz}, beyond their design program focused on precision Higgs and electrowek physics.  In the last decades, four projects of $\epem$ colliders have been proposed: the Compact Linear Collider (CLIC)~\cite{Aicheler:2012bya} and ILC linear machines, and the circular CEPC~\cite{CEPCStudyGroup:2018rmc} and FCC-ee facilities. In this paper, we present predictions for the range of c.m.\ energies and luminosities expected to be reached in various runs at the ILC and FCC-ee (Table~\ref{tab:runs}), but results for the other colliders can be obtained from the authors upon request. In our analysis, we consider that the colliding photons are emitted by the incoming $\rm e^\pm$ beams, although these initial photons can also be generated through Compton backscattering of laser photons  off linear electron beams~\cite{Telnov:1995hc}. Such an alternative has been explored in Ref.~\cite{Bose:2022obr}, which  demonstrated that the CLIC machine running at $\sqrts = 3000$~GeV, has a great physics potential for searching for ALPs with masses in the $m_a \approx 1$--2.4~TeV range. Our focus here is in the production of lighter ALPs ($m_a \approx 0.1$--1000~GeV) that can be produced in the range of collision energies listed in Table~\ref{tab:runs}.

\tabcolsep=1.9mm
\begin{table}[htbp!]
\centering
\caption{Center-of-mass energies and integrated luminosities expected in $\epem$ collisions during typical operation runs of the FCC-ee~\cite{FCC:2023} and ILC~\cite{ILD:2019kmq}.\label{tab:runs}}
\vspace{0.15cm}
\begin{tabular}{l|cccc}\hline
Collider/Detector\hspace{1cm} & \multicolumn{4}{c}{$(\sqrts,\LumiInt)$}\\
 & Run 1 & Run 2 & Run 3 & Run 4 \\\hline
FCC-ee/IDEA   & (91 GeV, 204~\abinv) & (160 GeV, 9.6~\abinv) & (240 GeV, 7.2~\abinv) & (365 GeV, 2.68~\abinv) \\
ILC/ILD      & (250 GeV, 2~\abinv) & (365 GeV, 0.2~\abinv) & (500 GeV,  4~\abinv) & (1 TeV, 5.4~\abinv) \\
\hline
\end{tabular}
\end{table}

The theoretical formalism needed to describe exclusive diphoton production in $\gaga$ interactions at $\epem$ colliders, represented in Fig.~\ref{Fig:diagram_epem} (center and left diagrams), is succinctly recalled. Such processes are characterized by  the presence of a pair of centrally produced photons and two forward $\epem$, mostly collinear to the incoming beams directions. Photon-photon collisions with higher virtuality lead to the electron and/or positron being deflected at increasingly larger angles with respect to the beamline. As demonstrated by the experimental analyses performed at the Large Electron-Positron Collider (LEP)~\cite{Przybycien:2008zz}, one or both scattered electrons can be detected, which allows focusing on higher-virtuality collisions and suppressing potential backgrounds at the cost of a reduction in the size of the data samples. The possibility of forward detector tagging at future $\epem$ colliders is under discussion. In our analysis, we do not require tagging of the electrons in the final state, and the inclusive process of interest is therefore dominated by low-virtuality $\gaga$ interactions where the EPA works best. The kinematics of the collision is described next. The incident particles emit two photons with momenta $q_1$ and $q_2$, which merge to form a diphoton system with invariant mass squared $m_{\gaga}^2 = (q_1 + q_2)^2 = (k_1 + k_2)^2$, where $k_1$ and $k_2$ are the momenta of the two photons in the final state. Following the EPA approach~\cite{Budnev:1975poe} for $\epem$ collisions at high energies ($\sqrts \gg m_e$), the two-photon production cross section can be written in the following general form
\begin{eqnarray}
    \sigma_{\epem}(\sqrts) = \frac{1}{2s} \int \frac{d^3p_1^{\prime}d^3p_2^{\prime}}{E_1^{\prime}E_2^{\prime}} \frac{d^3k_1}{2E_1 (2\pi)^3} \frac{d^3k_2}{2E_2 (2\pi)^3} \alpha(Q_1^2)\alpha(Q_2^2) \frac{\rho_1^{\mu \mu^{\prime}}\rho_2^{\nu \nu^{\prime}}M^*_{\mu^{\prime}\nu^{\prime}}M_{\mu\nu}}{Q_1^2Q_2^2} \times\delta(q_1 + q_2 - k_1 - k_2) \,\,,
    \label{sigma_epem}
\end{eqnarray}
where the outgoing electrons have momenta $p_{1,2}^{\prime}$ and energies $E_{1,2}^{\prime}$, $E_i$ are the energies of the photons in the final state, $Q_i^2 = - q_i^2$ are their virtualities, and $\alpha$ is the electromagnetic coupling constant. The matrix $\rho_i^{\mu\nu}$ can be interpreted as a density matrix for the virtual photon generated by the electron $i$, which can be fully estimated using quantum-electrodynamics for structureless pointlike particles such as electrons. In addition, $M^{\mu\nu}$ corresponds to the photon-photon amplitude, determined from all virtual charged particles running in the box of Fig.~\ref{Fig:diagram_epem} (right), with arbitrary initial-state photon virtualities. In our analysis, such amplitude is given by sum of the SM LbL contributions, including the W$^\pm$ boson and charged fermion loops~\cite{Bardin:2009gq}, plus the contribution associated to the presence of an ALP that couples to photons. 
The ALP contribution to exclusive diphoton production, represented by the $s$-channel fusion\footnote{In our analysis, we neglect the contributions from $t$ and $u$ channel ALP exchanges since they are very small and difficult to separate from other backgrounds.} shown the center panel of Fig.~\ref{Fig:diagram_epem}, is estimated considering the Feynman rules derived from the Lagrangian:
\begin{equation}
\mathcal{L} = \frac{1}{2}\partial^{\mu}\;a\;\partial_{\mu}a - \frac{1}{2}\;m_{a}^{2}\;a^{2} - \frac{1}{4} g_{a\gaga}\;a\;F^{\mu \nu}\,\Tilde{F}_{\mu \nu},    
\label{eq:lagrangian}
\end{equation}
where $m_{a}$ is the ALP mass, $g_{a\gaga}$ is the ALP-photon coupling and $(\Tilde{F}_{\mu \nu})\;F^{\mu \nu}$ is the (dual) electromagnetic field strength. The ALP diphoton width is given by
\begin{equation}
\Gamma_a = \frac{g_{a\gaga}^2m_a^3}{64 \pi}\,.
\label{eq:width}
\end{equation}
In what follows, we will assume that the ALP couples only to the photons, which implies a decay branching fraction of $\mathcal{B} = (a \to \gaga) = 100 \%$, and a total ALP width given by Eq.~(\ref{eq:width}). The ALP lifetime is given by the inverse of this equation, which for masses $m_a>1$~GeV and couplings $g_{a\gaga} > 0.1$~TeV$^{-1}$ compatible with the current experimental limits~\cite{dEnterria:2021ljz}, implies short-lived ALPs with transverse decay lengths smaller than $c\tau\ll 1$~mm, namely consistent with a diphoton system that is reconstructed as coming from the primary collision vertex. As we will discuss in Section~\ref{sec:results}, such an expectation is not automatically warranted for more weakly coupled ALPs ($g_{a\gaga} \ll 0.1$~TeV$^{-1}$), which are increasingly longer-lived and may appear as coming from a displaced secondary vertex or as missing energy in the event.\\


In order to estimate the ALP production rates and detection probabilities at $\epem$ colliders, using the formalism described above, we have generated events for the ALP signal and LbL continuum background with the SC4 Monte Carlo generator for all systems listed in Table~\ref{tab:runs}. 
The SC4 exclusive diphoton events 
are passed through the detector response simulated with the \delphes~3 code. For FCC-ee, we have used the latest IDEA detector setup, where photons are reconstructed in a dual readout calorimeter with  alternate clear and scintillating fibers embedded in a metal (lead, copper) matrix. Such an electromagnetic calorimeter (ECAL) can reconstruct photons with total energy $E_{\gamma} \geq 2.0$~GeV in the pseudorapidity region $|\eta|\leq 3.0$ with an efficiency of 99\%, where photon isolation/separation is defined within a cone aperture of 
$\Delta R = \sqrt{\Delta \phi^2 + \Delta \eta^2} > 0.5$ (where $\Delta \phi$ and $\Delta \eta$ are the azimuthal and pseudorapidity difference, respectively, of the photon with respect to any other reconstructed particle in the event) for $p_{\rm T\gamma}^\text{min}=0.5$~GeV. On the other hand, for the ILC case, we use the ILD detector parameters for a silicon-tungsten sampling ECAL where photons with total energy $E_{\gamma} \geq 2.0$~GeV in the pseudorapidity region $|\eta|\leq 3.0$ are detected with a 95\% efficiency, while for $3.0 < |\eta| \leq 4.0$ the efficiency is 90\%, maintaining the same photon isolation/separation settings as for the FCC-ee case. For future FCC-ee-related work, we aim to perform full simulation studies, including also the alternative CLD detector~\cite{Bacchetta:2019fmz}, and the newly proposed high-granularity noble liquid ECAL~\cite{Francois:2022apy} that is not yet available in the \delphes~3 package, using the turnkey software stack, Key4hep~\cite{Ganis:2021vgv}.

\section{Results}
\label{sec:results}

In the present study, the signal signature is characterized by two isolated photons exclusively produced, i.e.\ with no other activity in the detector, as well as no recoiling electrons detected.
The only SM background considered is that of the LbL continuum  because, as shown in~\cite{Steinberg:2021wbs} the potential $\epem \to \rm Z\gaga \to \nu_{\ell} \bar{\nu_{\ell}} \gaga$ background, with the neutrinos escaping undetected, can be well suppressed by requiring relatively low diphoton transverse momenta, $p_{T_{\gaga}}\leq 10$~GeV, without affecting the ALP signal yields.
In addition, the photon-fusion production of an $\epem$ pair, with both electron and positron further radiating each one of them a photon, $\gaga \to \epem \gaga$ (hard bremsstrahlung), will not result in an exclusive diphoton event, considering that the FCC-ee tracking detectors can reconstruct with full efficiency the tracks of the radiating $\epem$ pairs down to very small momenta~\cite{Barchetta:2021ibt}.
Since $\Gamma_a/m_a \ll 1$, the ALP signal appears as a narrow peak on top of the diphoton invariant mass $m_{\gaga}$ spectrum, defined as a function of the photon energies $E_1$, $E_2$ and their relative angle $\theta_{12}$: $m^2_{\gaga} = 2E_1 E_2 (1 - \cos \theta_{12})$.
In the region of masses $m_{\gaga}\approx 1$--5~GeV, there exist multiple (pseudo)scalar and tensor resonances that can be equally produced through photon fusion and that can decay to two photons~\cite{Shao:2022cly}. If the reconstructed (smeared) ALP mass coincides with the peak position of any such mesons, its observability would be jeopardized, and a more detailed resonance subtraction analysis would be needed in this particular mass region. Such a background-subtraction can be performed, as done e.g.\ for the search studies for true-tauonium diphoton decays at FCC-ee~\cite{dEnterria:2022ysg}, but it goes beyond the scope of this paper that aims at covering a much broader range of masses $m_a \approx 0.1$--1000~GeV.

\begin{figure}[htbp!]
\centering
\includegraphics[scale=0.25]{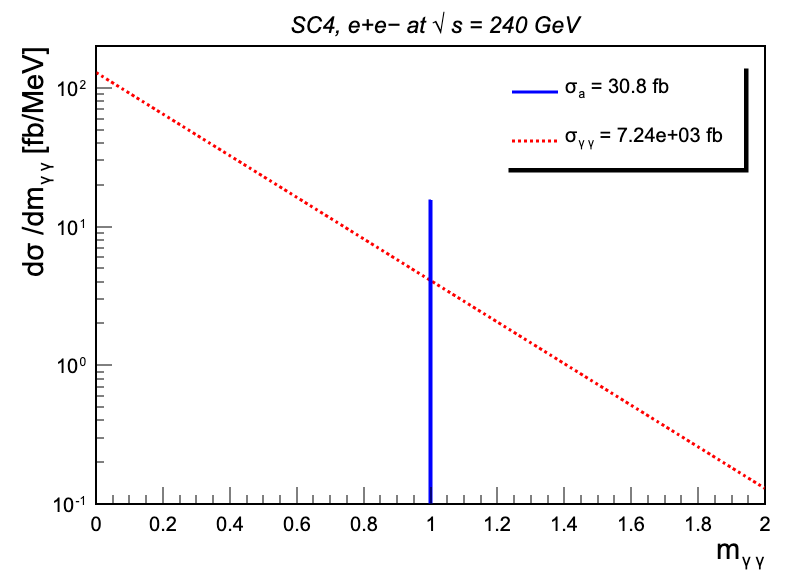}
\includegraphics[scale=0.25]{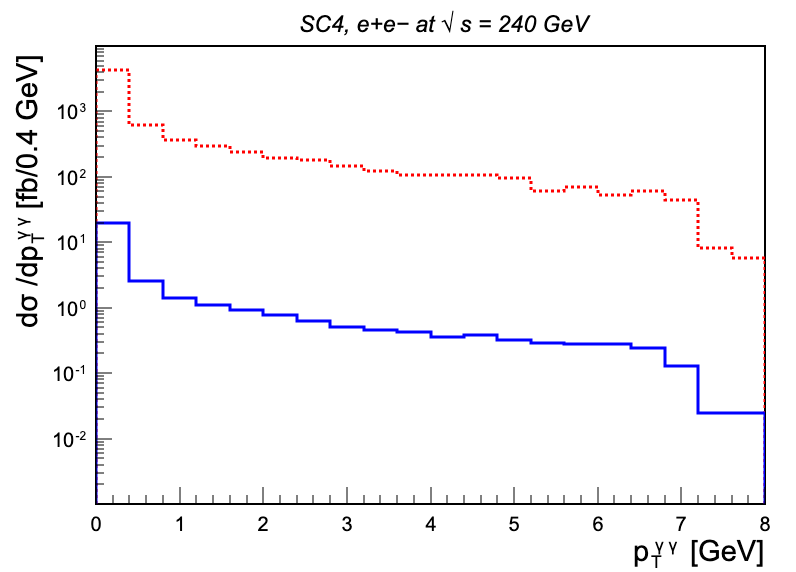}
\includegraphics[scale=0.25]{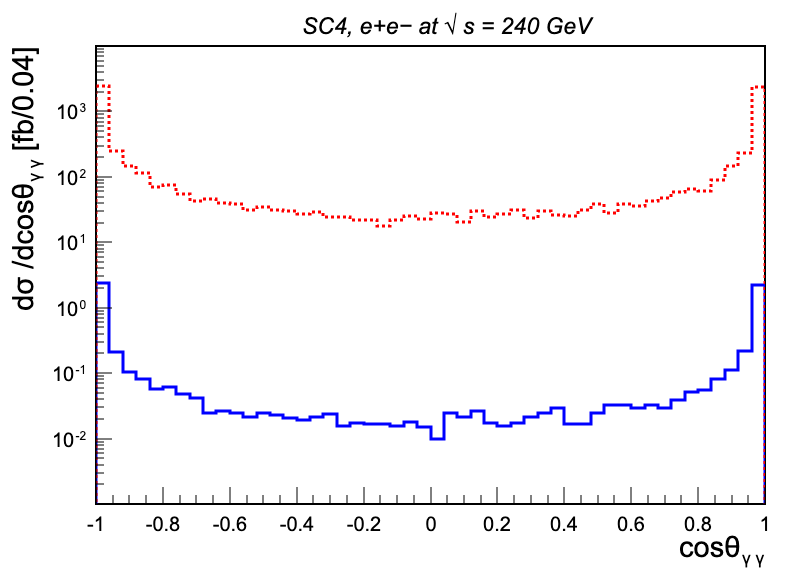}
\includegraphics[scale=0.25]{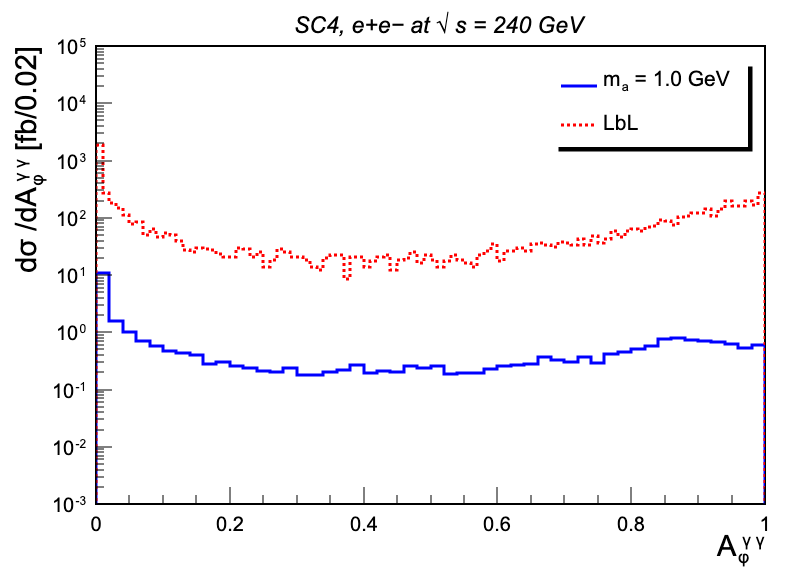}
    \caption{Photon pair generator-level kinematic distributions (invariant mass, transverse momentum, cosine polar angle, and acoplanarity) for an ALP with mass $m_{a} = 1.0$~GeV and  coupling $g_{a\gaga} = 0.1$~TeV$^{-1}$ (blue histogram) and for the LbL background around the same mass range (dashed red histogram) produced in $\epem$ collisions at $\sqrts = 240$~GeV. The diphoton invariant mass ALP peak has a width arbitrarily set to $\Gamma_a = 1$ MeV, to make it visible.
    \label{fig:240ALP1GeV}}
\end{figure}

\begin{figure}[htbp!]
\centering
     \includegraphics[scale=0.25]{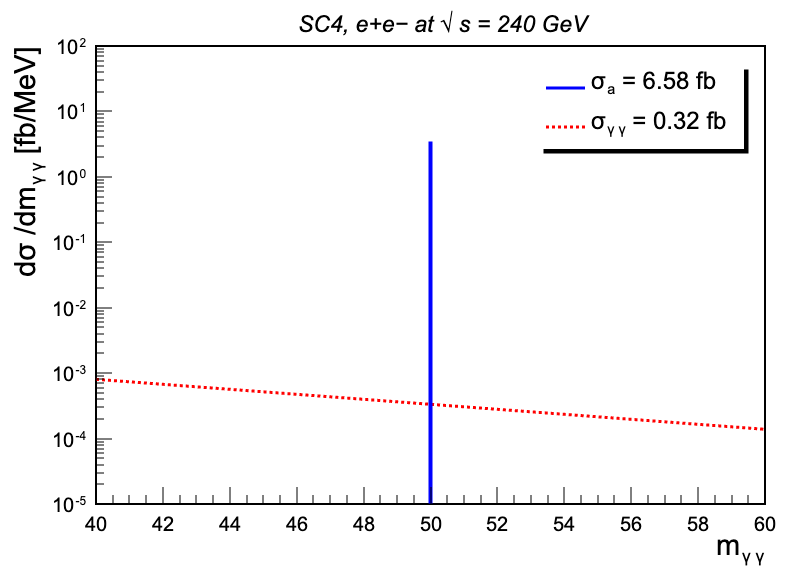}
     \includegraphics[scale=0.25]{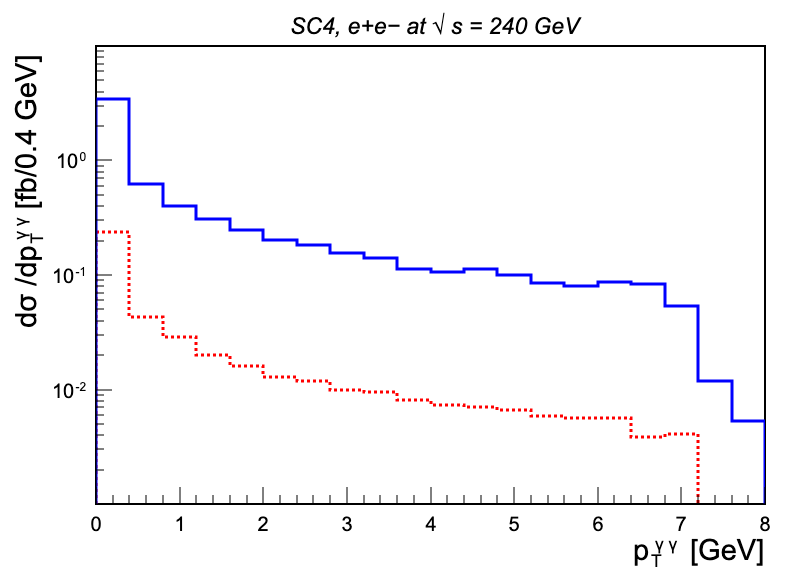}
     \includegraphics[scale=0.25]{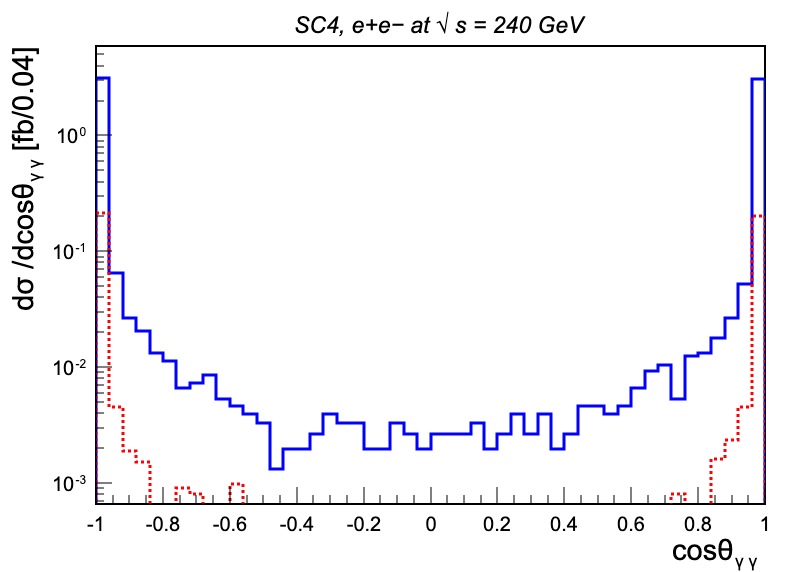}
     \includegraphics[scale=0.25]{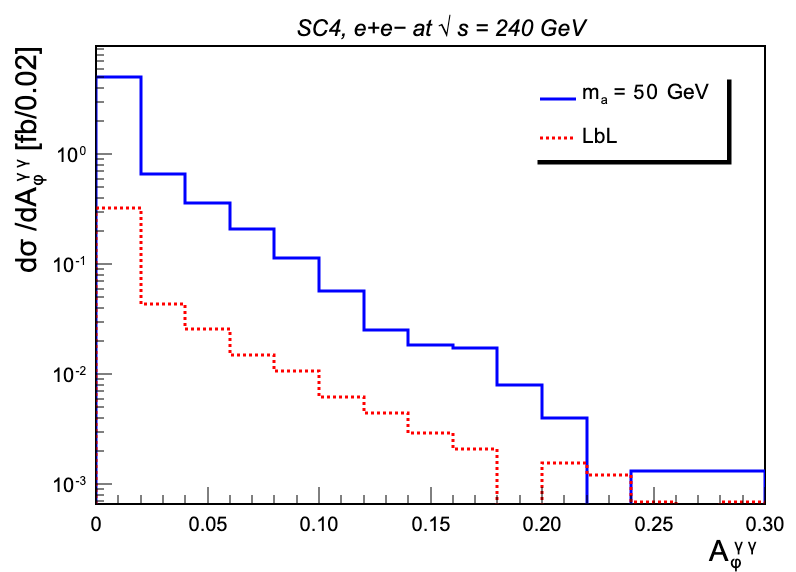}
    \caption{Photon pair generator-level kinematic distributions (invariant mass, transverse momentum, cosine polar angle, and acoplanarity) for an ALP with mass $m_{a} = 50$~GeV and  coupling $g_{a\gaga} = 0.1$~TeV$^{-1}$ (blue histogram) and for the LbL background around the same mass range (dashed red histogram)  produced in $\epem$ collisions at $\sqrts = 240$~GeV. The diphoton invariant mass ALP peak has a width arbitrarily set to $\Gamma_a = 1$ MeV, to make it visible.
    \label{fig:240ALP50GeV}}
\end{figure}

We initially select events with exactly two photons passing loose cuts in photon transverse momentum $p_{T\gamma} \geq 0.1$~GeV and pseudorapidity $|\eta_{\gamma}| \leq 3, 4$ (in the acceptance region covered by the IDEA and ILD electromagnetic calorimeters, respectively),
over a diphoton invariant mass $m_{\gaga}$ window around each assumed ALP mass. 
In Figs.~\ref{fig:240ALP1GeV} and~\ref{fig:240ALP50GeV}, the generator-level predictions for the diphoton invariant mass, transverse momentum, cosine of the polar angle with respect to the beam $\cos(\theta)$, and acoplanarity\footnote{Acoplanarity is defined as $A_\phi = A_\phi = 1 - |\phi_1-\phi_2|/\pi$ , where $\phi_{1,2}$ are the azimuthal angles of the final-state photons.} distributions are shown for $\epem$ collisions at $\sqrts = 240$~GeV and two values of ALP mass $m_a = 1.0\;\mbox{and}\;50$~GeV, respectively, for a fixed  $g_{a\gaga} = 0.1$~TeV$^{-1}$ coupling. 
The photon pair $p_\mathrm{T}$ and acoplanarity distributions peak at zero as expected for their production from the fusion of quasireal photons, leading to ALPs produced approximately at rest, but have relatively long tails due to the presence of $\gaga$ scatterings with larger virtualities. Obviously, the relevant kinematic distribution to identify the ALP signal is the $m_{\gaga}$ distribution, showing a resonance peak at the generated mass point, whereas all other ALP and LbL kinematic distributions are very similar in shape.

\begin{figure}[htbp!]
\centering  
\includegraphics[scale=0.25]{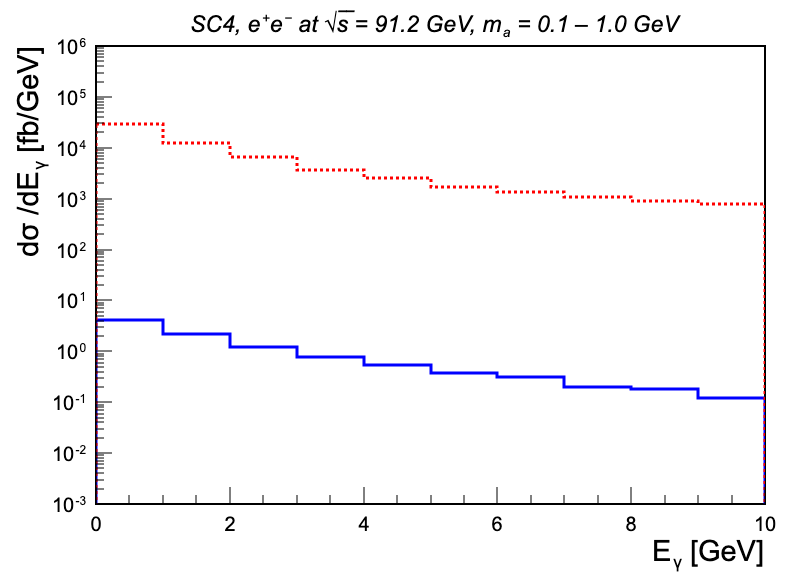}
\includegraphics[scale=0.25]{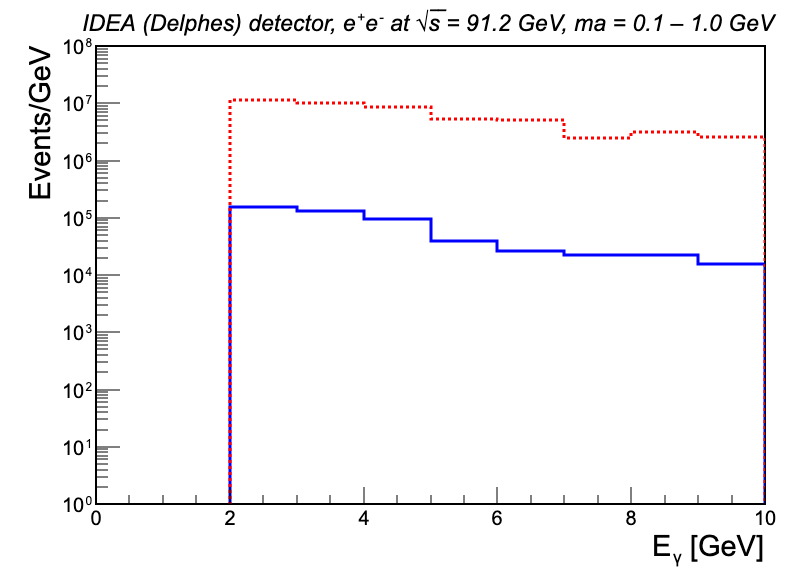}
\includegraphics[scale=0.25]{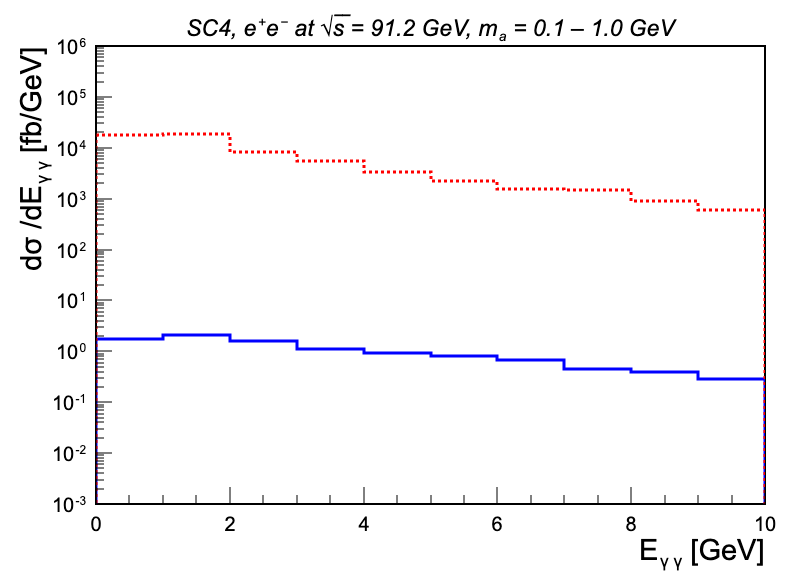}
\includegraphics[scale=0.25]{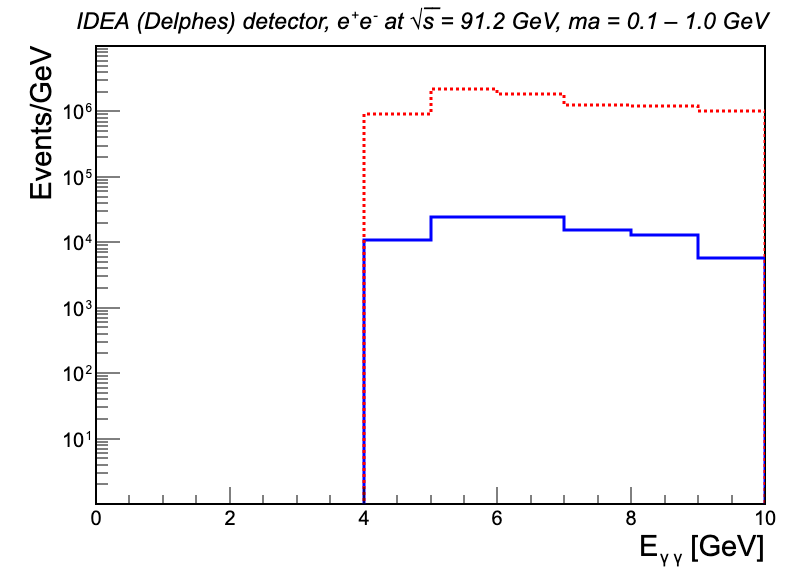}
    \caption{Distributions of the energy of single-$\gamma$ ($E_{\gamma}$) and $\gaga$ pairs ($E_{\gaga}$) for simulated ALPs in the mass range $m_{a} = 0.1$--1.0~GeV and coupling $g_{a\gaga} = 0.1$~TeV$^{-1}$ (blue histogram) and for the LbL continuum (dashed red histogram) in $\epem$ collisions at FCC-ee at $\sqrts =91.2$~GeV. The left (right) plots show generator-level (IDEA reconstruction-level) results. 
    \label{fig:energies}}
\end{figure}

\begin{figure}[!htbp]
\centering
\includegraphics[scale=0.3]{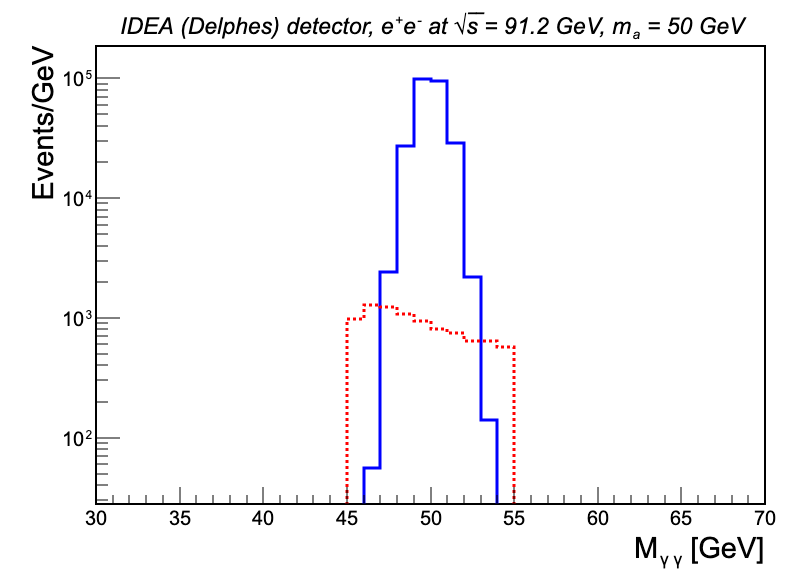}
    \caption{Invariant diphoton mass distribution for an ALP signal with mass $m_{a} = 50$~GeV and coupling $g_{a\gaga} = 0.1$~TeV$^{-1}$ (blue histogram) produced in $\epem$ collisions and reconstructed with the IDEA fast detector simulation at FCC-ee operating at $\sqrts= 91.2$~GeV, on top of the LbL continuum (dashed red histogram).
    \label{fig:Acopmasses}}
\end{figure}

The impact of the detector photon efficiency reconstruction can be seen in Fig.~\ref{fig:energies} for light ALPs with $m_a \approx 0.1$--1.0~GeV, where the left plots show the single-photon and diphoton energies at the input generator level, and the right plots show the same distributions after reconstruction with the parametrized detector response. The efficiency drops abruptly to zero for soft photons ($E_{\gamma} < 2$~GeV) in the IDEA \delphes~3 card, which considers a simplified response function at threshold. Although this has a large impact on the reconstruction efficiency for lighter ALP searches ($m_a < 1.0$~GeV), implementing a more realistic turn-on efficiency (or using alternative high-resolution crystal ECAL detectors being under consideration for at least one of the four interaction points at FCC-ee) requires a full detector simulation that goes beyond the scope of this paper.
Figure~\ref{fig:Acopmasses} shows the energy smearing effect on the reconstructed mass of an ALP with $m_a = 50$~GeV in the IDEA detector at FCC-ee operating at the Z pole. The prominent peak shown in Fig.~\ref{fig:240ALP50GeV} (top left) is washed out over several mass bins, thereby reducing the overall number of signal counts over a given $\Delta m_{\gaga}$ window, eventually (for increasingly small $g_{a\gaga}$ couplings) appearing just as an excess over the continuum background. A similar effect is of course present in the ILD detector reconstruction of diphotons (not plotted here). Mass smearing effects of this sort play a role in the reduction of the signal efficiency at all c.m.\ energies, but are comparatively less relevant for increasing ALP masses given the strong decrease of the underlying LbL cross section background as a function of $\gaga$ energy.\\

The photon-fusion searches for ALPs at hadron colliders have exploited the fact that the $\gamma$ fluxes are quasireal, and therefore that any produced resonances are basically at rest, and thus their decay diphotons are almost perfectly back-to-back in azimuth. In this case, a useful selection criterion to eliminate nonexclusive backgrounds is that of requiring a very small acoplanarity, $A_\phi < 0.01$~\cite{dEnterria:2013zqi,CMS:2018erd,ATLAS:2023zfc}. Such a criterion is less useful in $\epem$ collisions as the $A_\phi$ distributions have relatively long tails (Figs.~\ref{fig:240ALP1GeV} and~\ref{fig:240ALP50GeV}). For light ALPs with masses $m_a \lesssim 1.0$~GeV because the fusioning photons have virtualities of $\mathcal{O}(1$~GeV) and generate a final state with a transverse momentum of the same order, the two photons are clearly not back-to-back in azimuth. For higher ALP masses, the acoplanarity selection is also not useful because although the decay photons are increasingly back-to-back, the LbL continuum also shares the same kinematic property. As a matter of fact, for the lightest masses $m_a \lesssim 0.1$~GeV, the ALPs tend to decay into collimated photon pairs with inter-photon separation $\Delta R_{\gaga}\lesssim 0.2$, which essentially produce the same detector response as a single {\it fat photon} with the combined energy of both showers. The highly granular FCC-ee and ILC detectors will allow the identification of photons with very small separation from each other via a shower shape analysis. For the present analysis, we will not therefore apply any $A_\phi$ requirement.\\

The number of ALP signal events  ($s$) at a given diphoton mass $m_a$ and of LbL background events ($b$) in a window of $\Delta m_{\gaga}$ around it, obtained after applying the selection criteria on the reconstructed simulated data samples, are counted for each one of the experimental configurations listed in Table~\ref{tab:runs}. From the signal and background yields, the statistical significance $SS$ of the ALP signal over the background-only hypothesis (``bump hunting'') is computed using the maximum approximate median significance,
\begin{equation}
    SS = \sqrt{2\left( (s+b)\ln{\left(1+\frac{s}{b} \right)} -s \right)},
\end{equation}
for each proposed mass point and collider scenario (combination of c.m.\ energy and expected integrated luminosity). For $s \ll b$, one has that $SS$ reduces to $s/\sqrt{b}$. Systematic uncertainties are assumed to be small, and affect equally signal and background so they have no impact on the obtained $SS$. Finally, the parameter of interest (POI) is calculated using the asymptotic frequentist limits~\cite{Cowan:2010js} with the RootStats package~\cite{Moneta:2010pm}. The POI is the expected upper limit at 95\% confidence level (CL) on the signal strength $\mu$ obtained taking the ratio of $p$-values under the signal-plus-background and background-only hypotheses\footnote{For a detailed discussion we refer to the Section 5.1 in the reference~\cite{Cowan:2010js}.}.\\

As an example, Table~\ref{tab:FCC91} collects all relevant quantities obtained in this analysis for the FCC-ee run at the Z pole. For each ALP mass point, we list its theoretical cross sections for our reference $g_{a\gaga} = 0.1$~TeV$^{-1}$ coupling, the reconstruction efficiencies, and corresponding yields after cuts. The second column-block lists the same quantities for the LbL background in a window $\Delta m_{\gaga}$ around $m_a$. The third column-group of the table indicates the derived statistical significance, POI, and 95\% CL upper limits on the photon-ALP coupling and ALP production cross section.\\

\tabcolsep=2.mm
\begin{table}[htbp!]
\centering
\caption{Relevant numerical values for the statistical significance determination of an ALP signal (with $g_{a\gaga} = 0.1$~TeV$^{-1}$) on top of the LbL background in $\epem$ collisions at FCC-ee running at the Z pole. For each mass point $m_a$, we quote the ALP and LbL cross sections, selection criteria efficiency, and expected yields; as well as the statistical significance ($SS$), POI, and the derived 95\% CL upper limits on the photon-ALP coupling and ALP production cross section.\label{tab:FCC91}}
\vspace{0.15cm}
\resizebox{\textwidth}{!}{\begin{tabular}{cccc|cccc|ccccc}\hline
Mass      & ALP   & Detector & ALP  & $\Delta m_{\gaga}$ & LbL  & Detector & LbL & $SS$ & POI & $g_{a\gaga}$ @95\% & $\sigma$@95\% \\
{[}GeV{]} & $\sigma${[}fb{]} & efficiency & yield $s$ & [GeV] & $\sigma${[}fb{]} &  efficiency & yield  $b$ &  &     &{[}TeV$^{-1}${]} & [fb] \\
\hline
0.1 & 10.1  & 0.17 & $3.51\times 10^{5}$  & {[}0.05,0.15{]} & $5.98\times 10^{4}$ & 0.06 & $7.23\times 10^{8}$ & 13.1 & $1.50\times 10^{-1}$ & $3.87\times 10^{-2}$ & 1.52 \\
0.5& 26.9 & 0.06 & $3.32\times 10^{5}$  & {[}0.20,0.80{]} & $2.47\times 10^{4}$ & 0.03 & $1.70\times 10^{8}$ & 25.5 & $7.69\times 10^{-2}$ & $2.77\times 10^{-2}$ & 2.07  \\
1.0 & 25.3 & 0.11 & $5.81\times 10^{5}$ & {[}0.50,1.50{]}  & $5.97\times 10^{3}$ & 0.05 & $6.27\times 10^{7}$ & 73.3 & $2.67\times 10^{-2}$ & $1.63\times 10^{-2}$ & 0.67\\
10.0& 11.4 & 0.92 & $2.27\times 10^{6}$ & {[}8.0,12.0{]} & $2.37\times 10^{1}$ & 0.54 & $2.61\times 10^{6}$ & 1252.1 & $1.39\times 10^{-3}$ & $3.73\times 10^{-3}$ & $1.59\times 10^{-2}$  \\
50.0& 1.37 & 0.97 & $2.74\times 10^{5}$ & {[}45.0,55.0{]} & $7.7\times 10^{-2}$ & 0.61 & $9.64\times 10^{3}$ & 1170.6 & $7.06\times 10^{-4}$ & $2.66\times 10^{-3}$ & $9.71\times 10^{-4}$\\
85.0 & 0.13 & 0.93 & $2.58\times 10^{4}$ & {[}80.0,90.0{]} & $9.84\times 10^{-4}$ & 0.84 & $1.69\times 10^{2}$ & 458.1 & $1.04\times 10^{-3}$ & $3.22\times 10^{-3}$ & $1.40\times 10^{-4}$ \\\hline
\end{tabular}
}
\end{table}

\begin{figure}[!htbp]
\centering
\includegraphics[scale=0.33]{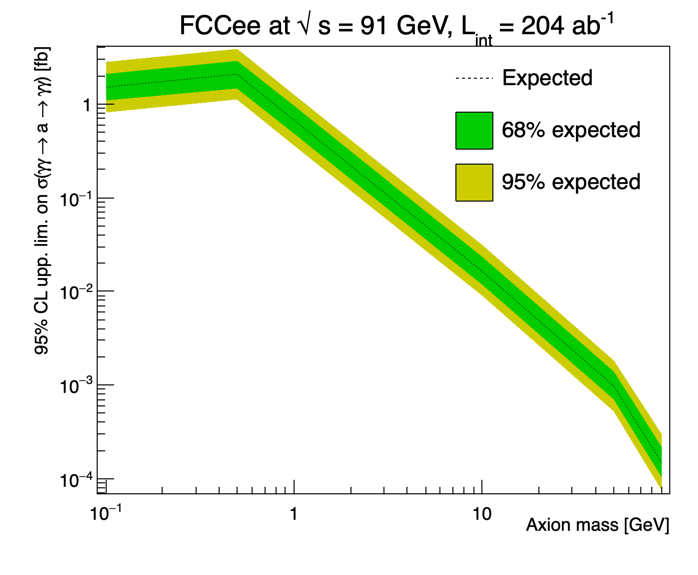}
\includegraphics[scale=0.33]{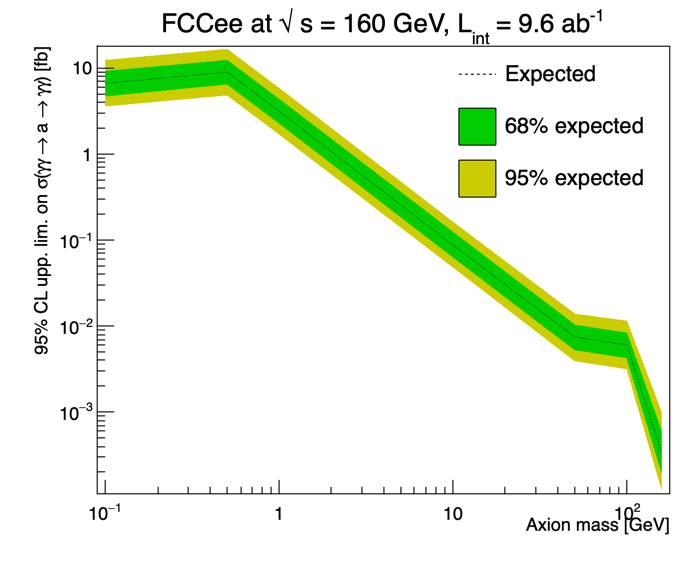}
\includegraphics[scale=0.33]{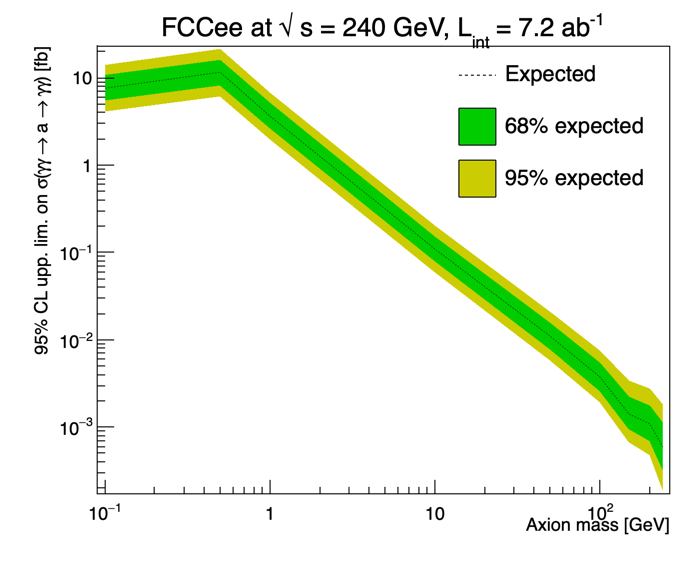}    
\includegraphics[scale=0.33]{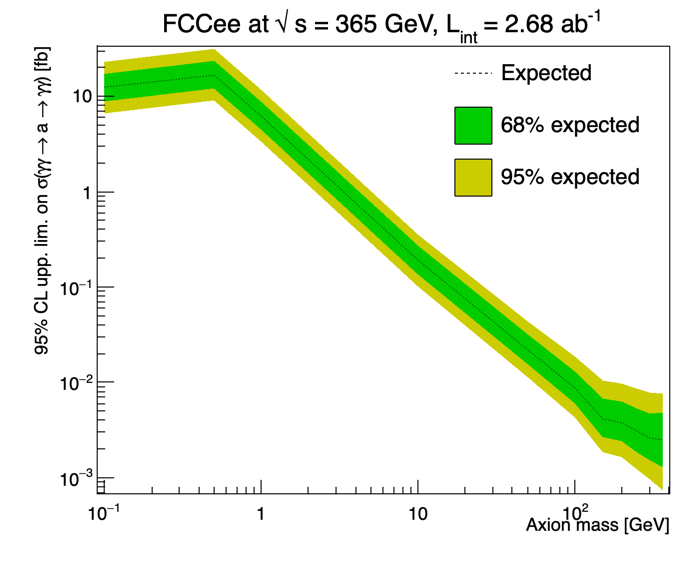}
    \caption{Expected upper limits (and 68 and 95\% CL bands) on the ALP production cross section $\sigma(\gaga \to a \to \gaga)$ as a function of ALP mass, reachable in the different FCC-ee runs (Table~\ref{tab:runs}).
    \label{fig:detectorXSFCCeeFull}}
\end{figure}
\begin{figure}[!htbp]
    \centering
    \includegraphics[scale=0.33]{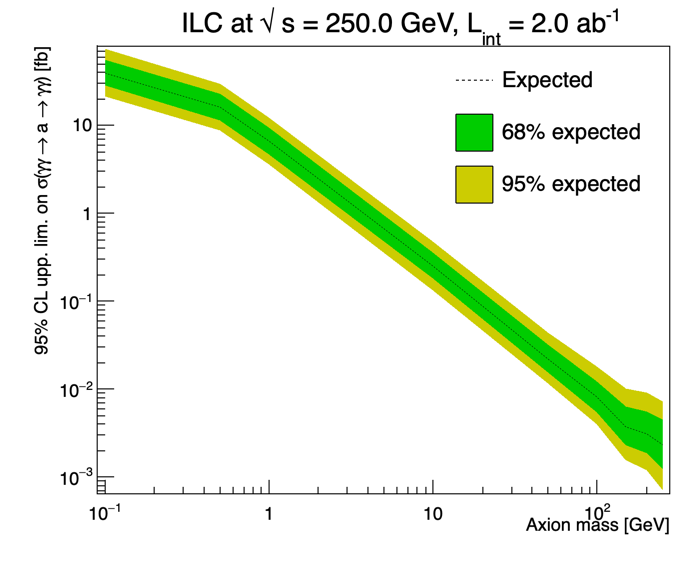}
    \includegraphics[scale=0.33]{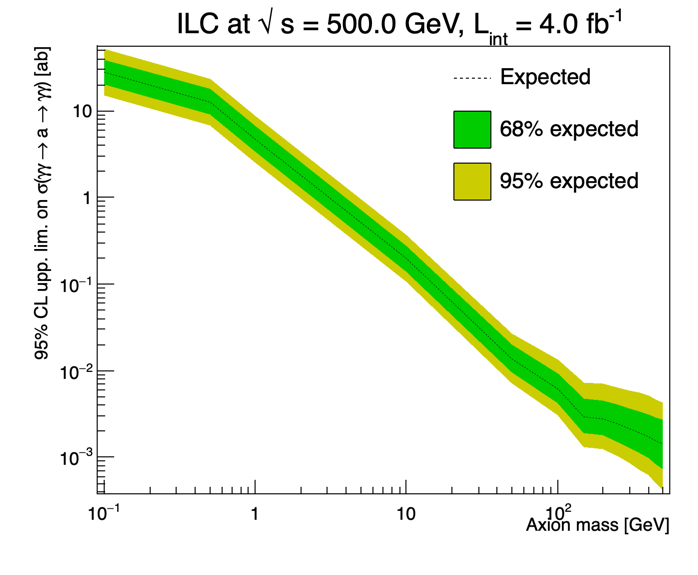}
   \includegraphics[scale=0.33]{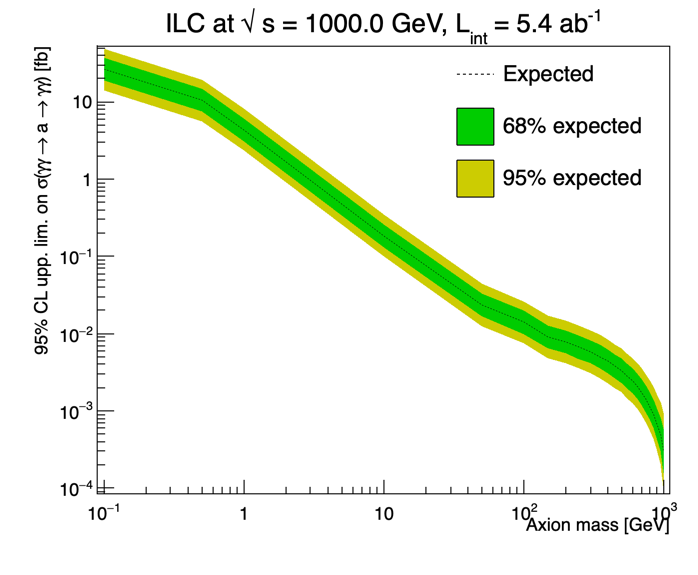}
    \caption{Expected upper limits (and 68 and 95\% CL bands) on the ALP production cross section $\sigma(\gaga \to a \to \gaga)$ as a function of ALP mass, reachable in the different ILC runs (Table~\ref{tab:runs}). The results for the run at 350~GeV are skipped given its comparatively lower integrated luminosity).
    \label{fig:detectorXSILCFull}}
\end{figure}

Figures~\ref{fig:detectorXSFCCeeFull} and~\ref{fig:detectorXSILCFull} present the expected limits of the ALP production cross section, and the 68\% and 95\% bands around it, for the ALP mass range $m_a \approx 0.1$--1000~GeV for the different $\epem$ collision runs expected at FCC-ee and ILC, respectively. The absence of any visible peak above the LbL continuum will allow the exclusion of ALP production cross sections above 1~(10)~fb at 95\% CL below $m_a \approx 1$~GeV at FCC-ee (ILC), decreasing down to $\sigma(\gaga\to a)\approx 1$~ab for $m_a \approx 300$~(1000)~GeV at FCC-ee (ILC).

\begin{figure}[!htbp]
\centering
\includegraphics[scale=0.35]
{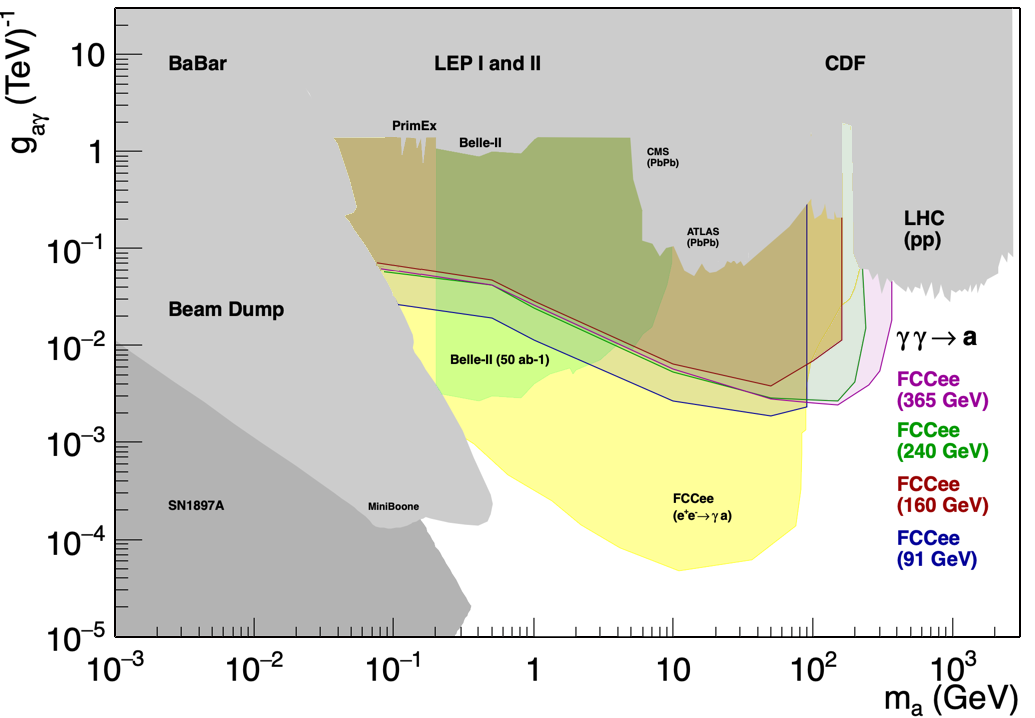}
\caption{Exclusion limits at 95\% CL on the ALP-photon coupling as a function of the ALP mass expected from searches for $\gaga\to a\to \gaga$ in the different FCC-ee runs (Table~\ref{tab:runs}). The yellow area shows FCC-ee expectations based on the alternative $\epem \to \gamma a$ final state~\cite{Bauer:2018uxu} scaled to reflect the updated FCC-ee operation. The green area shows current Belle-II upper bounds~\cite{Belle-II:2020jti,Dolan:2017osp} scaled up to the full expected SuperKEK integrated luminosity $\LumiInt = 50$~\abinv.
\label{fig:FCClimits4years}
}
\end{figure}

\begin{figure}[!htbp]
\centering
\includegraphics[scale=0.35]
{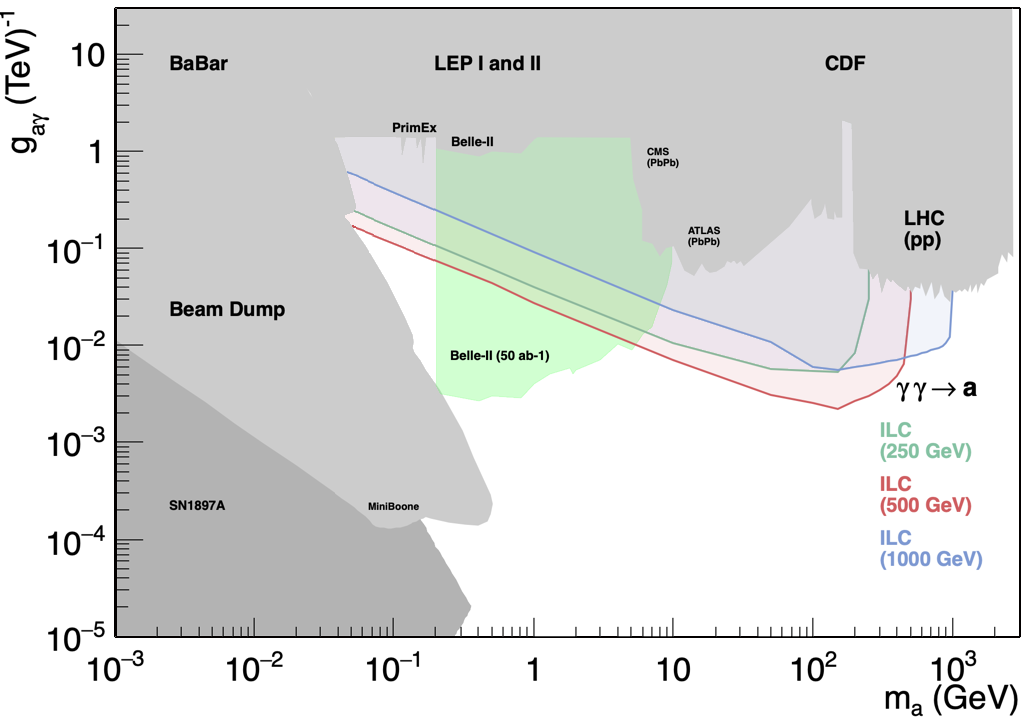}
\caption{Exclusion limits at 95\% CL on the ALP-photon coupling as a function of the ALP mass expected from searches for $\gaga\to a\to \gaga$ in the different ILC runs (Table~\ref{tab:runs}). The green area shows current Belle-II upper bounds~\cite{Belle-II:2020jti,Dolan:2017osp} scaled up to the full expected SuperKEK integrated luminosity $\LumiInt = 50$~\abinv.
\label{fig:ILClimits4years}}
\end{figure}

From the expected 95\% CL upper limits on the cross sections, one can derive the corresponding 95\% CL exclusion limits on the ALP-photon coupling as a function of the ALP mass that are shown in Fig.~\ref{fig:FCClimits4years} for FCC-ee and in Fig.~\ref{fig:ILClimits4years} for ILC. Our results are compared to current limits from $\epem$ and hadron colliders (including the latest bounds from p-p collisions~\cite{TOTEM:2021zxa,CMS:2022zfd,ATLAS:2023zfc}), as well as from beam-dump and astrophysical constraints~\cite{dEnterria:2021ljz,Antel:2023hkf}. Given that the significances are given by $SS \approx s/\sqrt{b}$, the individual curves for each $\epem$ run basically provide limits that comparatively scale according to the square-root of their relative luminosities given by Table~\ref{tab:runs}. Whereas, the large integrated luminosity of the FCC-ee run at the Z pole leads to a larger sensitivity in the range $m_a < 100$~GeV, the higher energy ILC runs allow probing the heavier $m_a \approx 350$--1000~GeV ALP range, which is otherwise inaccessible at FCC-ee. Both future $\epem$ factories will improve the current LHC limits by about one to two orders-of-magnitude over the mass range $m_a \approx 5$--1000~GeV, beyond what is statistically reachable in searches at the end of the HL-LHC. Figure~\ref{fig:FCClimits4years}  
also show  expectations based on the alternative $\epem \to \gamma a$ final state~\cite{Bauer:2018uxu}, scaled to reflect the updated FCC-ee operation~\cite{FCC:2023} (yellow area), as well as the current Belle-II upper bounds~\cite{Belle-II:2020jti,Dolan:2017osp} scaled up to the full integrated luminosity $\LumiInt = 50$~\abinv expected at SuperKEK~\cite{Belle-II:2010dht}. Below $m_a = m_\mathrm{Z}$, one can see that the $\epem \to \mathrm{Z} \to \gamma a$ search\footnote{We note that the work of~\cite{Bauer:2018uxu} ignored photon acceptance and efficiency effects, but those should be relatively small given that the ALPs from Z-boson decays are boosted and should produce final states with well-reconstructible energetic photons with $E_\gamma \approx 20$~GeV.}, which features virtually no background, will provide about two orders-of-magnitude better upper bounds than the $\gaga$ fusion production process considered here, but that the latter mechanism will be more competitive in the $m_a\approx 100$--350~GeV range. The limits over $m_a \approx 5$--2000~GeV expected at the end of the high-luminosity LHC (HL-LHC) phase, obtained simply by scaling the current bounds~\cite{CMS:2018erd,ATLAS:2020hii,TOTEM:2021zxa,CMS:2022zfd,ATLAS:2023zfc} by the square-root of the ratio of HL-LHC over currently exploited integrated luminosities, will be about a factor of ten better than the current LHC ones. Therefore, all in all, our results indicate that the FCC-ee will provide the best possible ALP sensitivity over $m_a \approx 0.1$--300~GeV. Such results exemplify the power of a future Tera-Z and Higgs factory such as FCC-ee, to search for weakly coupled BSM particles.\\

\begin{figure}[!htbp]
\centering
\includegraphics[scale=0.35]
{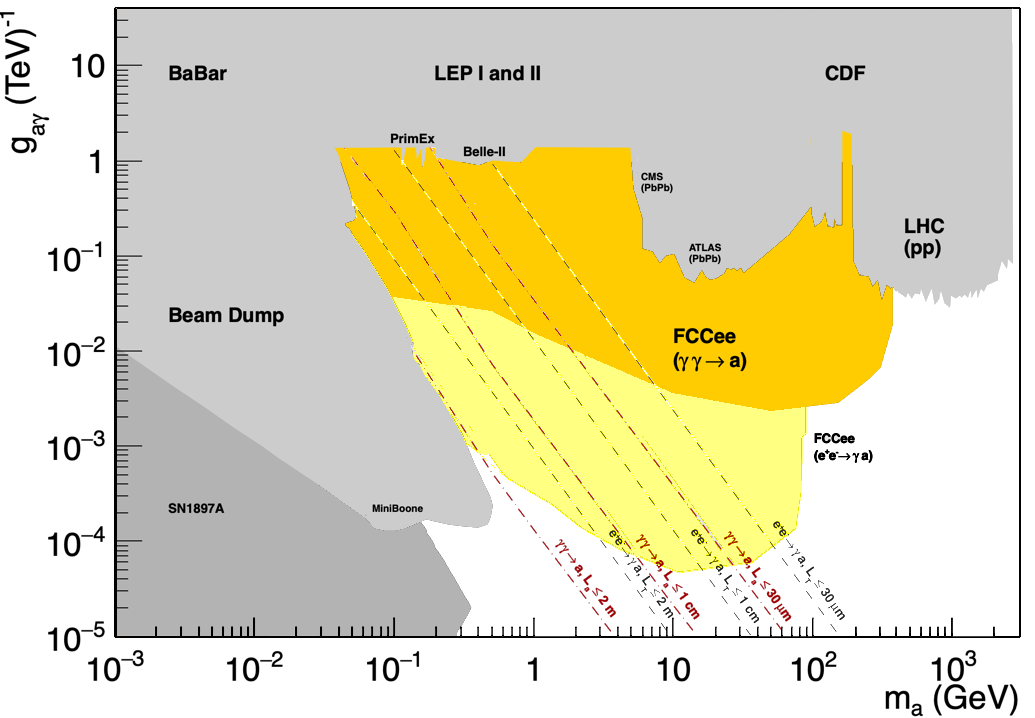}
\caption{Exclusion limits at 95\% CL on the ALP-photon coupling as a function of the ALP mass expected in $\epem$ collisions at FCC-ee for the combined $\gaga\to a$ (orange) and the $\epem\to a\gamma$ (yellow) processes, compared to current bounds (gray areas). Three reference average ALP transverse decay lengths, corresponding to $\langle L_{T} \rangle \approx 30~\mu$m,~1~cm,~2~m as per Eq.~(\ref{decaylength}), are indicated with dashed diagonal lines for both ALP production processes.
\label{fig:FCCCombined}}
\end{figure}

\begin{figure}[!htbp]
\centering
\includegraphics[scale=0.35]
{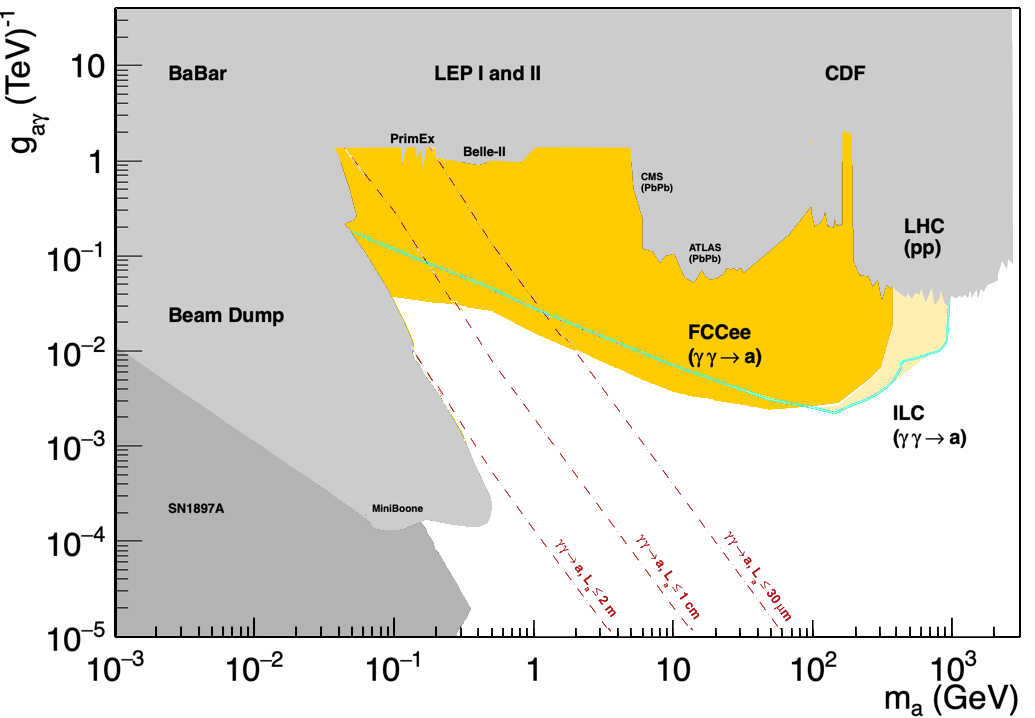}
\caption{Exclusion limits at 95\% CL on the ALP-photon coupling as a function of the ALP mass expected in all $\epem$ runs combined at ILC (beige) and at FCC-ee (orange) from the $\gaga\to a$ process, compared to current bounds (gray areas). Three reference average ALP transverse decay lengths, corresponding to $\langle L_{T} \rangle \approx 30~\mu$m,~1~cm,~2~m as per Eq.~(\ref{decaylength}), are indicated with dashed diagonal lines.
\label{fig:ILCCombined}}
\end{figure}

Light and weakly-coupled ALPs are long-lived, and their finite decay length can be used to reduce prompt backgrounds (coming from the primary event vertex) in experimental searches at colliders. The basic formula, in natural units ($\hbar = c = 1$), for the decay length in the rest frame of the ALP is $L_{a} = 1/\Gamma_a$, where the width $\Gamma_a$ is given by Eq.~(\ref{eq:width}). In the laboratory (collider) frame, we need to take into account the relativistic boost of the ALP, in which case $L$ is related to the rest-frame decay length by $L=\gamma \beta L_{a}$, where $\gamma$ and $\beta$ are the standard Lorentz factors. 
However, the decay length $L$ computed in this way is along the path of the ALP, and we are interested in the transverse plane (i.e., perpendicular to the beam) direction where the collider experiments have displaced vertexing capabilities. Given the angle $\theta$ between the ALP's direction and the beam axis,
the transverse decay length can be obtained from the expression $L_{T}=L\sin{\theta}$. Putting everything together, the average ALP decay length in the transverse direction reads
\begin{equation}
   \langle L_{T}\rangle = \langle \gamma \beta\rangle\, L_{a}\sin{\theta} =  \langle(\gamma \beta)_{T}\rangle\,L_{a} = 64 \pi \; \langle p_{T}\rangle \; g_{\gaga}^{-2} \; m_a^{-4},
\label{decaylength}
\end{equation}
where $p_T, g_{\gaga}, m_a$ are given in GeV, and one uses the $\hbar c = 0.197\cdot10^{-15}$~m\,GeV$^{-1}$ conversion factor to get the proper $L_T$ units in meters.
Being the average decay length proportional to $g_{a\gaga}^{-2}$ and to $m_{a}^{-4}$, the smaller the ALP mass and coupling are, the more long-lived the ALP will be. If boosted enough, the ALP decay may appear as a displaced vertex, or even beyond the detector volume manifesting itself as missing transverse energy in the event. 
For indicative purposes, in Figs.~\ref{fig:FCCCombined} and~\ref{fig:ILCCombined} we show curves in the $(m_a,g_\mathrm{a\gaga})$ plane corresponding to average ALP transverse decay lengths of $\langle L_T\rangle = 30~\mu\mbox{m},$~1~cm, and 2~m. The choice of these three baseline lengths is driven by experimental considerations: For diphoton final states, a secondary vertex can be determined for $L_T \gtrsim 1$~cm using ECAL pointing capabilities (namely, using the fine ECAL granularity to orient the combined towers to align with the original $\gamma$ direction), or for $L_T \gtrsim 30~\mu$m if both photons suffer an $\epem$ conversion, and the detector fiducial volume in the transverse direction, beyond which the ALP would be an invisible particle, covers a radius of $L_T \approx 2$~m. Depending on the ALP boost, i.e.\ on the $\langle p_T \rangle$ induced in its production process, the ALP will appear  separated enough from the primary $\epem$ interaction vertex, so that its displaced decay can be isolated, or not. The ALP search based on the triphoton decay of the Z boson, $\epem \to Z(\gamma a) \to 3\gamma$ (Fig. \ref{Fig:diagram_epem}, left), has very small SM backgrounds~\cite{dEnterria:2023wjq} and is the most competitive one in the $m_a<m_\mathrm{Z}$ region. In this case, the ALP is boosted by a large transverse momentum $\langle p_T\rangle \approx m_\mathrm{Z}/2 \approx 45$~GeV. On the other hand, in the photon-fusion process the ALP boost in the transverse direction is just given by the (small) virtualities of the colliding $\gamma$'s and amounts to $\langle p_T\rangle \approx 1$--3~GeV according to our simulations. Figure~\ref{fig:FCCCombined} compares the $\epem\to a\gamma$ (beige) and $\gaga\to a$ (orange) limits expected at FCC-ee, with their three corresponding average ALP transverse decay length ranges: $\langle L_{T} \rangle \approx 30~\mu$m,~1~cm,~2~m as per Eq.~(\ref{decaylength}), indicated with dashed diagonal lines. In the region of limits reachable at FCC-ee, one can see that most of the ALPs from the $\gaga$ production mode will have an average decay length below 1~cm and, thus, will be indistinguishable from the primary vertex, whereas a significant fraction of those coming from the $\epem \to Z(\gamma a)$ decay for the lowest $g_{a\gaga}$ couplings will feature secondary vertices within $\langle L_T \rangle \gtrsim 1.0$~cm~(1.0~m) for $m_a\lesssim 10$ (1.0)~GeV.


Finally, Fig.~\ref{fig:ILCCombined} compares the ALPs limits in the $(m_a,g_{a\gaga})$ plane expected via $\gamma\gamma\to a$ from all runs combined at the ILC (beige) and at FCC-ee (orange). Three reference average ALP transverse decay lengths $\langle L_{T} \rangle \approx 30~\mu$m,~1~cm,~2~m as per Eq.~(\ref{decaylength}) are indicated with dashed diagonal lines. For most of the phase space covered, the ALPs will appear as coming from the primary vertex.

\section{Summary}
\label{sec:conclusions}

We have presented feasibility studies for the observation of axion-like particles (ALPs) produced via photon-fusion processes and decaying in the diphoton mode, $\gaga \to a \to \gaga$, in $\epem$ collisions at the FCC-ee and ILC future colliders. Parametrized simulations of the photon response corresponding to two types of detectors (IDEA and ILD) at both colliders are used to evaluate the impact of the $\gamma$ acceptance and reconstruction efficiencies. Event selection criteria are applied aiming at identifying a resonant diphoton excess on top of the light-by-light continuum background. The full analysis, from hard scattering to the detector simulation, is performed for different ALP mass points, and for c.m.\ energies and integrated luminosities corresponding to four typical operation runs at FCC-ee and ILC, showing the achievable sensitivity on the photon-induced ALP production and on its potential discovery. 
Upper limits at 95\% confidence level on the cross section for ALP production, $\sigma(\gaga \to a \to \gaga)$, and on the ALP-photon coupling are obtained over the $m_a \approx 0.1$--1000~GeV mass range, and compared to current and future collider searches.
The FCC-ee operation at the Z pole, thanks to its enormous integrated luminosity, offers an exceptional opportunity to discover ALPs within the mass window $m_a \approx 0.1$--91.2~GeV, as well as to achieve the most stringent limits on the axion-photon coupling down to $g_{a\gaga}\approx 2\cdot 10^{-3}\;\mbox{TeV}^{-1}$ via $\gaga$ fusion, or even a factor of thirty better, down to $g_{a\gaga}\approx 6\cdot 10^{-5}\;\mbox{TeV}^{-1}$, using the alternative $\epem \to Z \to a\gamma \to 3\gamma$ channel with very small SM backgrounds. The higher c.m.\ energies of the ILC runs allow probing the heavier $m_a \approx 350$--1000~GeV range, which is otherwise inaccessible at FCC-ee. 
Over the ALP mass range $m_{a}\approx 5$--1000~GeV, both FCC-ee and ILC will supersede current (and foreaseable future) limits set at the LHC on the axion-photon coupling for all running scenarios, emphasizing the important role that such future $\epem$ facilities will play on searches for new weakly coupled particles.


\begin{acknowledgments}
P.R.T. warmly thanks the support of FAPERJ (grants no. E-26/201.816/2020, E-26/200.598/2022, and 210.362/2022) and the CMS Collaboration at CERN. D.E.M. thanks dearly, in the person of Janusz Chwastowski, the total support of the Henryk Niewodniczanski Institute of Nuclear Physics Polish Academy of Sciences (grant no. UMO2021/43/P/ST2/02279). V.P.G. was partially supported by CNPq, CAPES, FAPERGS and  INCT-FNA (Process No. 464898/2014-5) 

\end{acknowledgments}

\bibliographystyle{myutphys}
\bibliography{axionbibliography}

\end{document}